\documentclass[10pt,twocolumn,letterpaper]{article}

\usepackage{iccv}
\usepackage{times}
\usepackage{epsfig}
\usepackage{graphicx}
\usepackage{amsmath}
\usepackage{amssymb}

\usepackage{commath}

\usepackage{multirow}
\usepackage{multicol}
\usepackage{float}
\usepackage{comment}
\usepackage{wasysym}
\usepackage{comment}

\usepackage{subfig}
\usepackage[mathscr]{euscript}

\usepackage[usenames,dvipsnames]{color} 

\newcommand\SpaceReduction{-0.31 cm}

\usepackage[pagebackref=true,breaklinks=true,letterpaper=true,colorlinks,bookmarks=false]{hyperref}

\iccvfinalcopy 

\def\iccvPaperID{10655} 

\ificcvfinal\fi

\begin{document}

\title{4D Feet: Registering Walking Foot Shapes Using Attention Enhanced Dynamic-Synchronized Graph Convolutional LSTM Network}

\author{Farzam Tajdari$^1$, Toon Huysmans$^2$, Xinhe Yao$^3$, Jun Xu$^4$, Yu Song$^5$\\
Technical University of Delft \\
{\tt\small \{f.tajdari, t.huysmans, x.yao-1, j.xu-6, y.song\}@tudelft.nl}
}

\maketitle
\ificcvfinal\thispagestyle{empty}\fi

\begin{abstract}
   4D scans of dynamic deformable human body parts help researchers have a better understanding of spatiotemporal features. However, reconstructing 4D scans based on multiple asynchronous cameras encounters two main challenges: 1) finding the dynamic correspondences among different frames captured by each camera at the timestamps of the camera in terms of dynamic feature recognition, and 2) reconstructing 3D shapes from the combined point clouds captured by different cameras at asynchronous timestamps in terms of multi-view fusion. In this paper, we introduce a generic framework that is able to 1) find and align dynamic features in the 3D scans captured by each camera using the nonrigid iterative closest-farthest points algorithm; 2) synchronize scans captured by asynchronous cameras through a novel ADGC-LSTM-based network, which is capable of aligning 3D scans captured by different cameras to the timeline of a specific camera; and 3) register a high-quality template to synchronized scans at each timestamp to form a high-quality 3D mesh model using a non-rigid registration method. With a newly developed 4D foot scanner, we validate the framework and create the first open-access data-set, namely the 4D feet. It includes 4D shapes (15 fps) of the right and left feet of 58 participants (116 feet in total, including 5147 3D frames), covering significant phases of the gait cycle. The results demonstrate the effectiveness of the proposed framework, especially in synchronizing asynchronous 4D scans using the proposed ADGC-LSTM network. 
\end{abstract}

\vspace{\SpaceReduction}
\vspace{-0.3 cm}
\section{Introduction}
\label{sec:intro}

Human movements often lead to large shape deformation of different body parts. 4D scanning, which is able to capture 3D geometric shapes over time, attracted a lot of attention for a better understanding of the dynamic anthropometry in different applications \cite{tajdari2022next, tajdari2021image, boorady2009protective, bye2006analysis}, as 4D scans help researchers establish the fundamental understanding of human movements and give valuable information regarding human body deformation while performing different types of activities. Outcomes of research on 4D scans can be applied in many areas, e.g. building virtual avatars, performing (virtual) ergonomics evaluations, developing computer games, designing personal protective equipment, workwear, sportswear, and other practical garments \cite{boorady2011functional, Minnoye2022PERSONALIZED}. 
\begin{figure}[t]
	\centering
	\includegraphics[width= 1\linewidth]{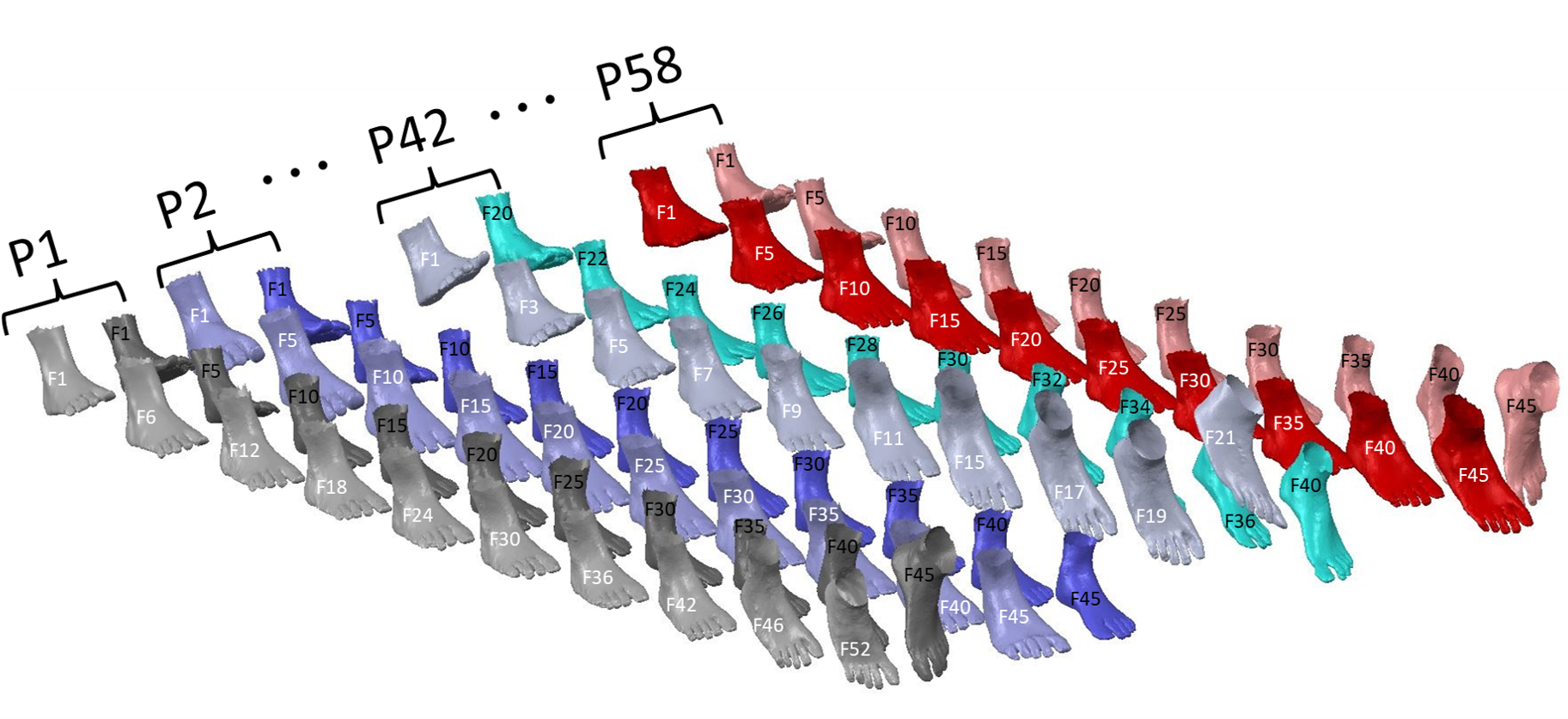}
 \vspace{-0.5 cm}
	\caption{4D Feet. We present a new 4D data-set of 58 Participants (P1..P58 in the figure), including 5147 frames of 3D scans. The raw 3D scans (meshes) were collected at 15 fps through a novel 4D foot scanner including 6 Azure Kinect DK cameras. Then we showed how to synchronize the cameras through a novel deep-learning-based framework, and establish a mesh-morphed data-set. 
 }
	\label{fig:Our_Data_set}
	\vspace{-0.43 cm}
        \vspace{-0.43 cm}
\end{figure}

To acquire 4D scans, multiple (depth) cameras are often used. Those cameras can be synchronized for capturing continuous images at a given moment. However, it is difficult to balance the needed resolutions of the images, the needed time duration, the buffer of the depth cameras, the data transfer rate, the computing power, and the storage~\cite{blackwell202154}. For instance, to capture 640$\times$576 depth images by 6 cameras at 30 frames per second (fps), the needed bandwidth is about 2Gb/s. This poses challenges in the design of a 4D scanning system, especially for a low-cost system. "Dropped frames" are frequently observed in the captured data, \textcolor{black}{mainly due to that the huge amount of to-be-transferred data leads to a nonlinear accumulative delay regarding each camera, even if all cameras are hardware-based synchronized~\cite{AzureKinectDelay}. In a practical case of using 6 Azure Kinect DK cameras for 15 fps 4D scanning, as in Figure~\ref{fig:Scanner}, even when all cameras are hardware synchronized, we found that there are on average 2 milliseconds delays for each frame acquired by those cameras in a 3-seconds scanning session. Note that the delay is accumulative, i.e. at the beginning of the scanning all the cameras' outputs are well-aligned based on their clocks; however, the longer the duration of the scanning is, the more the delay accrues, resulting in a divergence of the geometry in each frame regarding the timestamps.}


Researchers developed different methods, on both hardware and software, to tackle those challenges. In the scanning of a human body (parts), a possible approach is to use the prior knowledge of human actions and the associated dynamic features to synchronize the captured frames. In the past decades, recognizing human dynamics features has attracted a lot of attention in the field of computer vision. The developed 3D human action recognition methods can be roughly classified as the RGB video-based approaches \cite{wang2016temporal, wang2018rgb}, skeleton-based methods \cite{shi2019skeleton, si2019attention}, depth image-based methods \cite{yang2014super, yang2016super, xiao2019action} and the point cloud-based method \cite{wang20203dv}. Although the existing methods are proven to be effective in many applications, e.g., video surveillance, human-computer interaction, sports analysis \cite{9795869, poppe2010survey}, most of them are limited to employ (depth) images as the input, and the recognized 3D actions as the output. Extracting point-to-point correspondences among sequential point clouds from multiple views e.g., cameras, is rarely investigated \cite{wang20203dv, qi2017pointnet++, bilen2017action, liu2021geometrymotion}. On this topic, there are two fundamental challenges: 1) establishing the dynamic connectivity among asynchronous images (scans) captured by different cameras in terms of dynamic feature synchronization i.e. temporal correspondence, and 2) extracting meaningful dynamic features from the combined camera views for accurate analysis of deformation, i.e. multi-view fusion.

In this paper, using a newly developed low-cost 4D foot scanner based on 6 Microsoft Azure Kinect DK depth cameras, we developed a framework to synchronize and register the captured asynchronous images on significant phases of the gait cycle, resulting in a new open-access 4D Feet data-set of 58 subjects (116 feet). Our main contributions are:
\begin{itemize}
    \item Establishing and implementing a simple and effective framework to synchronized spatiotemporal asynchronous scans captured from multiple cameras and to track a point's correspondences in all the frames to extract dynamic features of each vertex e.g, velocity; 
    \item Developing a novel Attention Enhanced Dynamic-Synchronised Graph Convolutional (ADGC)-LSTM network to synchronize the dynamic features extracted from different cameras besides existing algorithms;
    \item Presenting the first 4D mesh-morphed walking foot open-access data-set (4D Feet), as a validation of the proposed framework. 
\end{itemize}

\section{Related work}

\subsection{Skeleton-and-depth-based action recognition}

The skeleton-based approach and the depth-based approach are often used in recognizing dynamic features of human actions based on prior knowledge ~\cite{liu2021geometrymotion}. Regarding skeleton-based 3D action recognition, sequence-based approaches, and graph-based approaches are often used. Via describing the skeleton as a sequence of joints, the sequence-based approaches~\cite{shi2019skeleton, si2019attention} employed the RNN (Recurrent Neural Network) based methods to extract temporal connectivity among those featured points. The graph-based approaches~\cite{li2019actional, zhang2019view} often utilized GCN (Graph Convolution Network) to exploit spatiotemporal connectivity by considering the skeletal structure as a graph, where the featured points are considered as the points of the graph. Regarding depth-based 3D action recognition, the available methods~\cite{oreifej2013hon4d, yang2014super, ohn2013joint, wang2018depth} mainly use the visualization features through 2.5D depth maps. Although both approaches are able to give a reasonably good estimation of the 3D actions that the target subject performed, it is difficult to form a generic framework to fully extract dynamic features based on a few featured points of the moving object, which may cause a reduction in the performance of 3D action recognition.

\subsection{3D Point Clouds action recognition}

Deep learning tools play a key role in extracting human actions via 3D point clouds, which is widely employed in recent studies \cite{qi2017pointnet++, liu2019relation, li2018so, su2018splatnet, liu2019flownet3d, gu2019hplflownet, liu2019meteornet, behley2019semantickitti, tajdari2020intelligent}. Among them, PointNet \cite{qi2017pointnet} as one of the most recent methods in the area, employs a set of MLPs on each of the individual vertices to identify the unique features. Next, it utilizes a max-pooling layer to generate the global identifier for each point cloud which does not use any geometry-based connectivity of the local neighboring structure. Contrary to these single-frame-based point cloud analysis approaches, in this paper, we present a simple and effective framework for time series 3D shape reconstruction and action recognition, in which we explicitly use temporal information in the motion stream to identify dynamic features. 

\vspace{\SpaceReduction}
\section{Methodology}
\label{sec:Methodology}
In this section, we first present the newly developed 4D foot scanner which is used to capture 4D data. Then, details of the proposed novel ADGC-LSTM network are presented. Based on the hardware and the novel ADGC-LSTM network, we introduce the framework for synchronizing asynchronous 4D scans from different cameras. Finally, we introduce the mesh registration method to along synchronized scans of different cameras at each timestamp.  

\subsection{The 4D foot Scanner}
\label{subsec:Scanner}
\vspace{-0.2 cm}

\begin{figure}[tb]
	\centering
	\includegraphics[height= 0.5\linewidth]{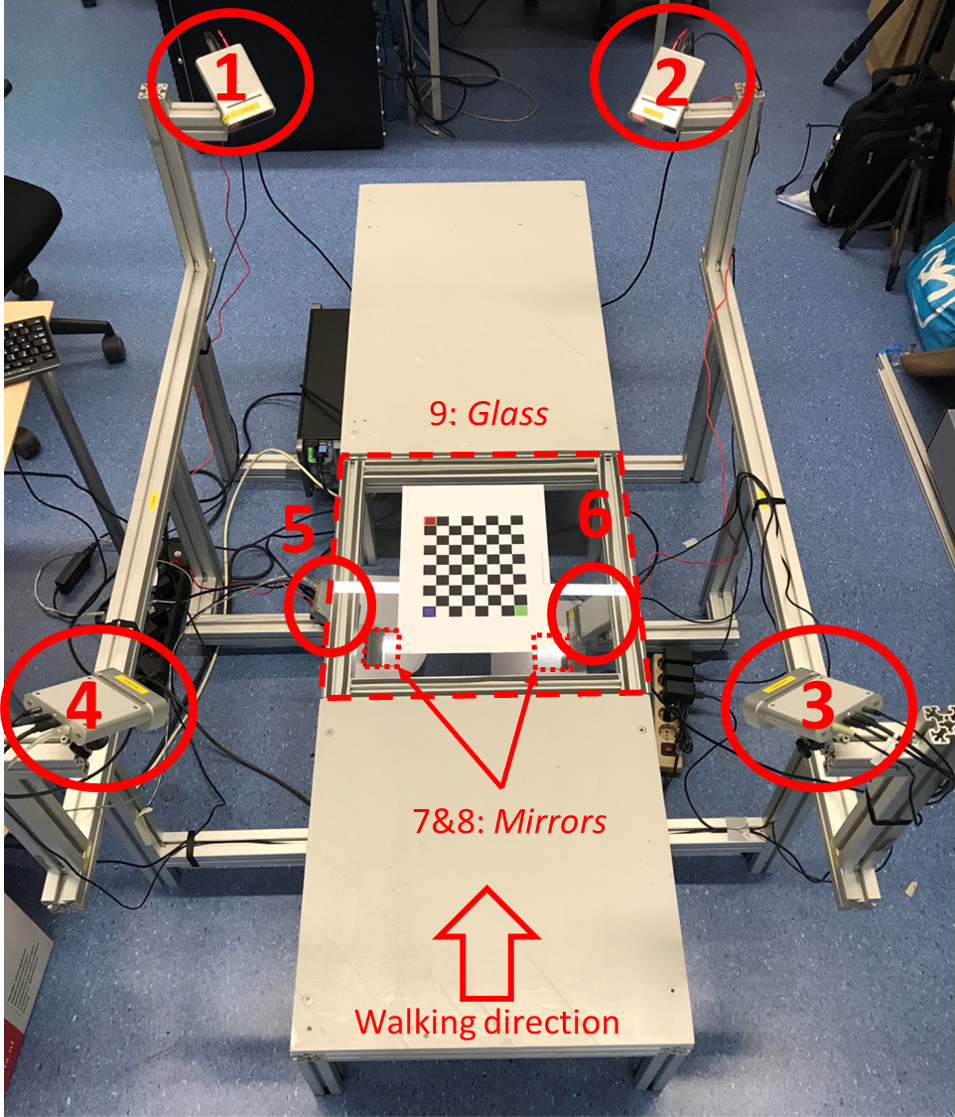}
 \vspace{-0.2 cm}
	\caption{The TU Delft 4D foot Scanner.}
	\label{fig:Scanner}
	\vspace{-0.43 cm}
\end{figure}

A 4D foot scanner was developed at TU Delft \cite{Kwa2021DesignPodiatrists, Tajdari2022OptimalPosition} for acquiring dynamic foot shape data. Figure~\ref{fig:Scanner} presents the next generation of the 4D foot scanner which utilizes six Microsoft Azure Kinect DK  cameras to capture the 4D foot shapes, where four cameras are installed on the top (id 1, 2, 3, and 4) and two cameras are at the bottom (id 5, and 6). To adapt to the minimal focal distance( $\sim{50}$ cm) of the cameras at the bottom, two first-surface mirrors (id 7 and 8) were placed on the floor to "fold" the optical path for   lowering the height of the scanner for a better user experience. A 9 mm thickness plexglass (id 9) was installed on the footpath to enable capturing the shape of the bottom of the foot while a subject is walking. 

The spatial positions and orientations of all cameras were optimized to maximize the resolutions of the captured scans and the intersections of effective view volumes of 6 cameras \cite{Kwa2021DesignPodiatrists}. To transform the captured data to a global coordinate system, we used a two-sided checkerboard shown in Figure~\ref{fig:Scanner}, and the code in \cite{geiger2012automatic} is utilized.

\subsection{Attention Enhanced Dynamic-synchronised Graph Convolutional LSTM network}
\label{subsec:DSG}

In the analysis of sequential geometric shapes, many  studies suggested that the LSTM, as a transformation of RNN, has a strong capability to understand long-term time dependency of the phenomenons e.g., understanding temporal dynamics of limited points-network (skeleton) sequences. However, using LSTM alone is difficult for incorporating spatial relations in the limited points-network-based action recognition. To this end, AGC-LSTM~\cite{si2019attention}, as an extension of LSTM, was developed to incorporate not only unique features of spatial configuration and temporal dynamics but also the coincident relationships between the spatial domain and temporal domain.  

In the process of capturing moving objects (4D scanning), the requirements of the needed movement ranges and the limited views of the cameras are always contradictory factors. This often results in a compromise in the design of the 4D scanner, either with a very small working envelope with limited movements or with sparse points in some of the captured point clouds. Commercial systems may employ more camera modules in 4D scanning; however, at the cost of investment and increased complexity. As the human movements are part of nature and cannot be constrained in a limited range, we target at building correspondences between/among sparse point clouds. Therefore the principles of the graph convolution model which has been broadly employed in sequential data with limited points-network nodes were adopted. Establishing the graph model plays a fundamental role in the graph convolution algorithm. Available graph convolution models e.g, AGC-LSTM, have several limitations for example using single graph structures, illed-correspondences among points, and inadequate discrimination of dissimilar actions. Here we develop a graphic model according to the Dynamic-Synchronised Graph based on the dynamic points, aiming at generating more sparse dynamic features to enhance the capability of the AGC-LSTM model in classifying spatiotemporal features and improve the precision of the action recognitions. The proposed novel method presented here is named Attention Enhanced Dynamic-Synchronized Graph Convolutional LSTM Network (ADGC-LSTM). The details of the algorithm are presented below.

\begin{figure}[tb]
	\centering
	\includegraphics[width= \linewidth]{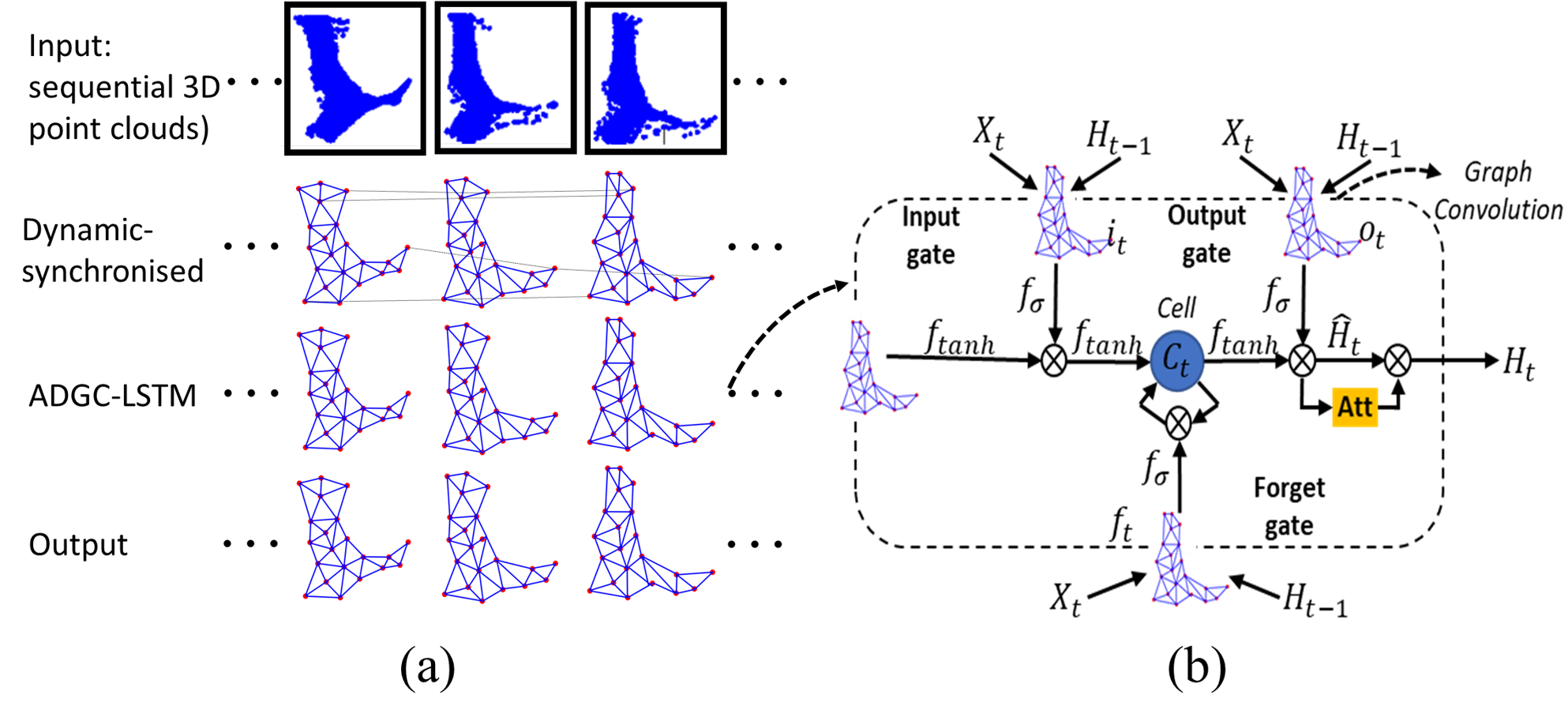}
 \vspace{-0.2 cm}
	\caption{The structure. a) One ADGC-LSTM layer; b) One ADGC-LSTM unit adapted from \cite{si2019attention}.}
	\label{fig:Structure}
	\vspace{-0.43 cm}
\end{figure}

Following the structure of LSTM, the ADGC-LSTM includes three gates: the input gate $i_t$, forgetting gate $f_t$, and output gate $o_t$. The input $X_t$, hidden state $H_t$, and cell memory $C_t$ are graph structure data, and the graph structure is generated by the ICFP (Nonrigid Iterative Closest-Farthest Points) algorithm explained in Section~\ref{subsubsec:Construction}. The graph convolution operator in the ADGC-LSTM, cell memory $C_t$, and hidden state $H_t$ can be used to extract temporal dynamics, and include spatial structure information. Figure~\ref{fig:Structure}(a) describes the structure of an ADGC-LSTM layer. Figure~\ref{fig:Structure}(b) describes the structure of the ADGC-LSTM unit. Equation \eqref{eq:LSTMgates} describes the functions of the ADGC-LSTM unit.\vspace{\SpaceReduction}
\begin{equation}
\begin{split}
    i_t &= \sigma \left( W_{xi} \star \textsl{g} X_t + W_{hi} \star \textsl{g} H_{t-1} + b_i\right) \\
    f_t &= \sigma \left( W_{xf} \star \textsl{g} X_t + W_{hf} \star \textsl{g} H_{t-1} + b_f\right) \\
    o_t &= \sigma \left( W_{xo} \star \textsl{g} X_t + W_{ho} \star \textsl{g} H_{t-1} + b_o\right) \\
    u_t &= \textrm{tanh} \left( W_{xc} \star \textsl{g} X_t + W_{hc} \star \textsl{g} H_{t-1} + b_c\right) \\
    C_{t} &= f_{t} \odot C_{t-1} + i_{t} \odot u_{t}\\
    \hat{H}_{t} &= o_t \odot \textrm{tanh}(C_t)\\
    H_t &= f_{att}(\hat{H}_t) + \hat{H}_t
    \end{split}
    \label{eq:LSTMgates}
    \vspace{\SpaceReduction}
\end{equation}
where $\star \textsl{g}$ defines the graph convolution operator and $\odot$ defines the Hadamard product. $\sigma (.)$ denotes the sigmoid activation function. $u_t$ denotes the modulated input. $\hat{H}_t$ explains an intermediate hidden state. $W_{xi} \star \textsl{g} X_t$ defines a graph convolution of $X_t$ with $W_{xi}$. The used graph convolution is the same as the graph convolution employed for the Graph Convolutional Neural (GCN) network in~\cite{yan2018spatial} with $K$ number of labels.
 $f_{att}(.)$ is an attention network that can select the diverse information of key nodes. The output $H_t$reinforces the information of key nodes, without neglecting the information of non-focus nodes, aiming at better integrity of spatial information.

\begin{figure}[tb]
	\centering
	\includegraphics[width= 0.5\linewidth]{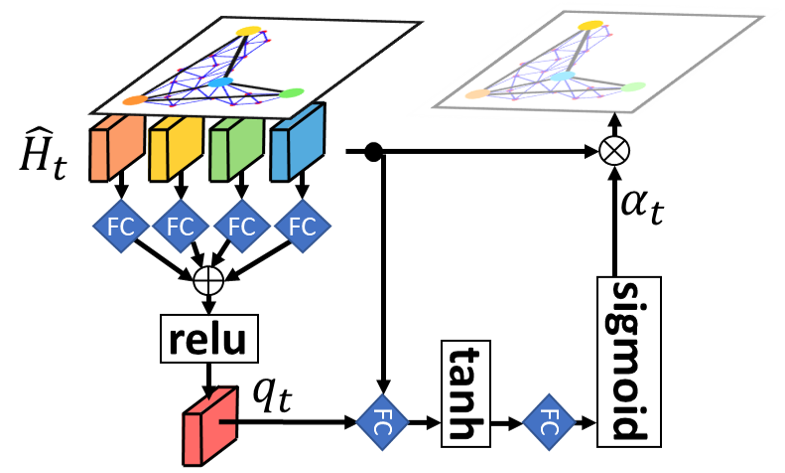}
 \vspace{-0.2 cm}
	\caption{Illustration of spatial attention mechanism principal adapted from \cite{si2019attention}.}
	\label{fig:Fig3}
	\vspace{-0.43 cm}
\end{figure}

\color{black}
The ADGC-LSTM network logically insists on key nodes by using a soft attention mechanism that automatically quantifies the emphasis level of the key nodes. The importance of the spatial attention network is depicted in Figure~\ref{fig:Fig3}. The intermediate hidden state ($\hat{H}_t$) of ADGC-LSTM contains persistent spatial structure information and temporal dynamics. The state practically improves the selection of the key nodes procedure. In order to guarantee that independent degree weights are established and reinforce the significance of dissimilar nodes for dissimilar types of actions, we employed a query feature as:

\vspace{\SpaceReduction}\begin{equation}
    q_t = relu \left( \sum_{i = 1}^{N} W\hat{H}_{t_i} \right)
\end{equation}
where $W$ defines the trainable parameter matrix, and $N$ is the number of nodes in the graph. Thus the attention scores of all nodes would be specified as:
\vspace{\SpaceReduction}\begin{equation}
    \alpha_t = sigmoid \left(\!\! U_s\;tanh \left(\! W_h \hat{H}_t \!+\! W_q q_t \!+\! b_s  \!\right) + b_u \!\!\right)
\end{equation}
where $\alpha_t = (\alpha_{t_1}, \alpha_{t_2}, \dots, \alpha_{t_N})$, and $U_s$, $W_h$, and $W_q$ are the trainable matrices. $b_s$ and $b_u$ are the bias. A non-linear function $sigmoid$ is employed regarding the probability of selected key joints. The hidden state $H_{t_i}$ of node $v_{t_i}$ is considered as $(1+\alpha_{t_i})\hat{H}_{t_i}$. The attention
enhanced hidden state $H_t$ is considered as an input for the next ADGC-LSTM layer. In the final layer of the ADGC-LSTM network, the
accumulation of all node features is classified as a global feature $F_t^g$, and the weighted sum of focused nodes is classified as a local feature $F_t^l$:\vspace{\SpaceReduction}
\begin{equation}
    F_t^g = \sum_{i=1}^{N} H_{t_i}; \;\; F_t^l = \sum_{i=1}^{N} \alpha_{t_i} \hat{H}_{t_i}.
    \vspace{\SpaceReduction}
\end{equation}
\color{black}
\begin{figure}[tb]
	\centering
	\includegraphics[width= 0.7\linewidth, height = 1.0\linewidth ]{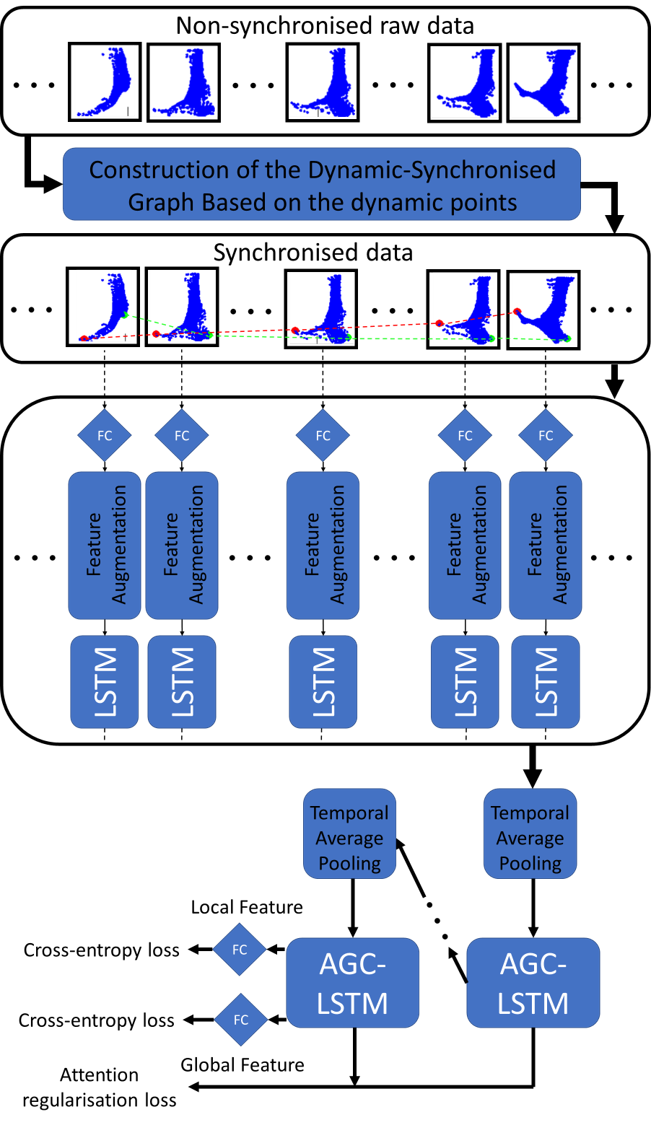}
 \vspace{-0.2 cm}
	\caption{Time synchronization procedure.}
	\label{fig:Fig6}
	\vspace{-0.43 cm}
\end{figure}

\subsection{Time synchronization}
\label{subsec:estimator}

 In Figure~\ref{fig:Fig6}, the workflow of the proposed time synchronization method is presented based on the data captured by the 4D scanner and the proposed novel ADGC-LSTM network.
In the following, Section~\ref{subsubsec:Construction} focuses on explaining the process of synchronizing data for each camera as the first step in the workflow. Then in Section~\ref{subsubsec:ADGC_LSTM_Network}, the aligned data from all cameras is synchronized using the proposed ADGC-LSTM network. Section~\ref{subsubsection:Implementation_ADGC_LSTM} describes details in the implementation of ADGC-LSTM regarding scans from different cameras.

\vspace{\SpaceReduction}
\subsubsection{Construction of the Dynamic-Synchronised Graph Based on the dynamic points}
\label{subsubsec:Construction}

\begin{figure}[tb]
		\centering
    \subfloat[Camera 1]{\includegraphics[width = 0.8\linewidth, height = 0.17\linewidth]{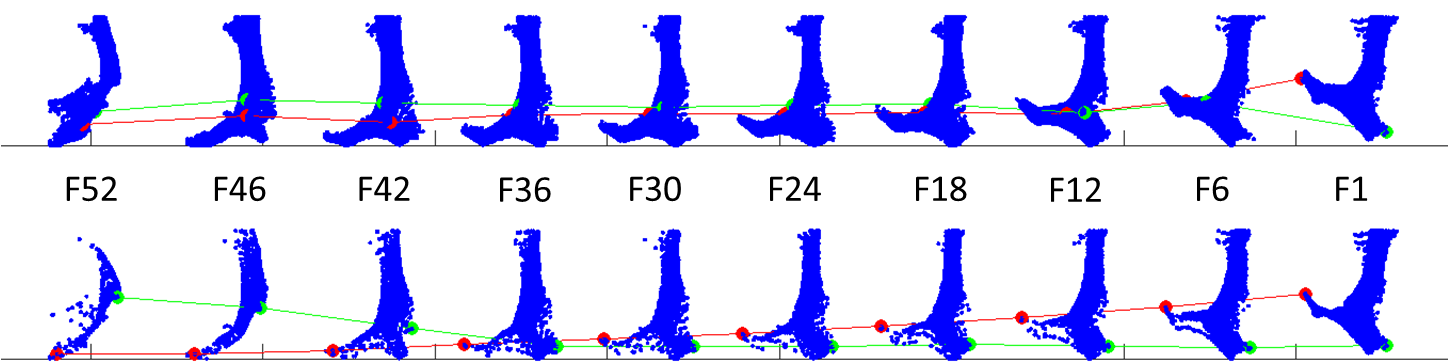}} \vspace{-0.4 cm}\\
    \subfloat[Camera 2]{\includegraphics[width = 0.8\linewidth, height = 0.17\linewidth]{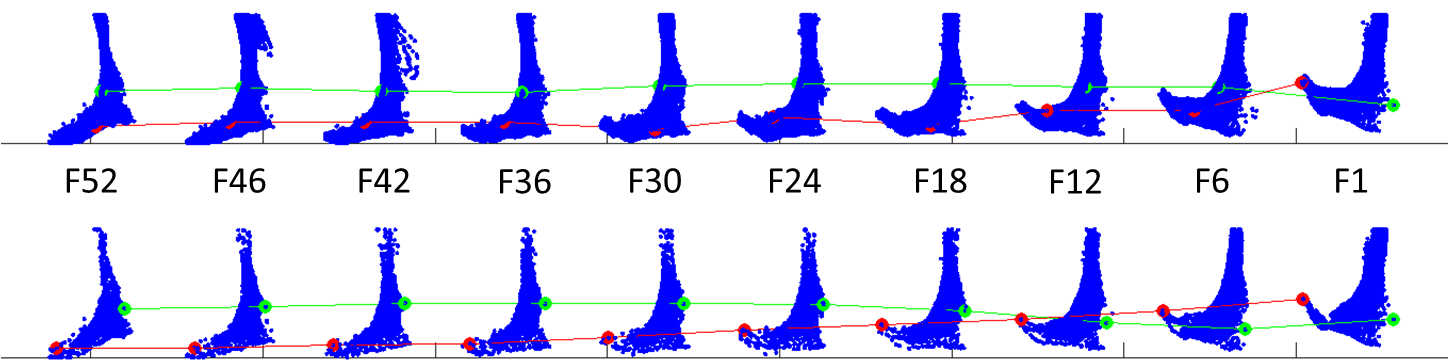}} \vspace{-0.4 cm}\\
    
    \subfloat[Camera 3]{\includegraphics[width =  0.8\linewidth, height = 0.17\linewidth]{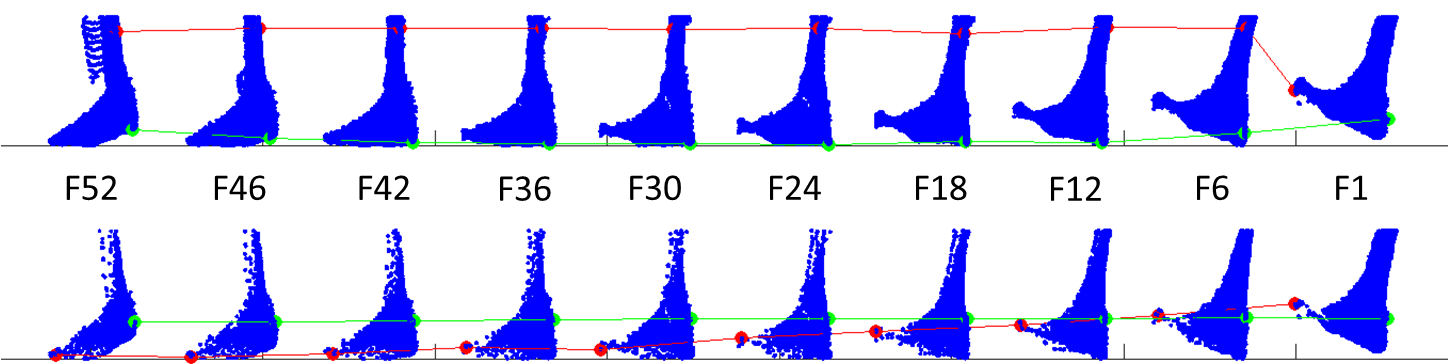}} \vspace{-0.4 cm}\\
    \subfloat[Camera 4]{\includegraphics[width =  0.8\linewidth, height = 0.17\linewidth]{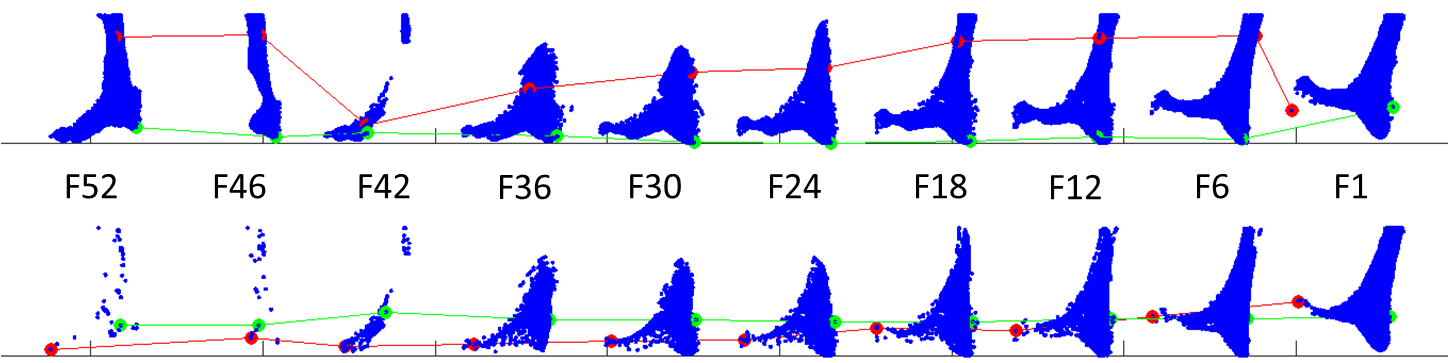}} \vspace{-0.4 cm}\\
    
    \subfloat[Camera 5]{\includegraphics[width =  0.8\linewidth, height = 0.17\linewidth]{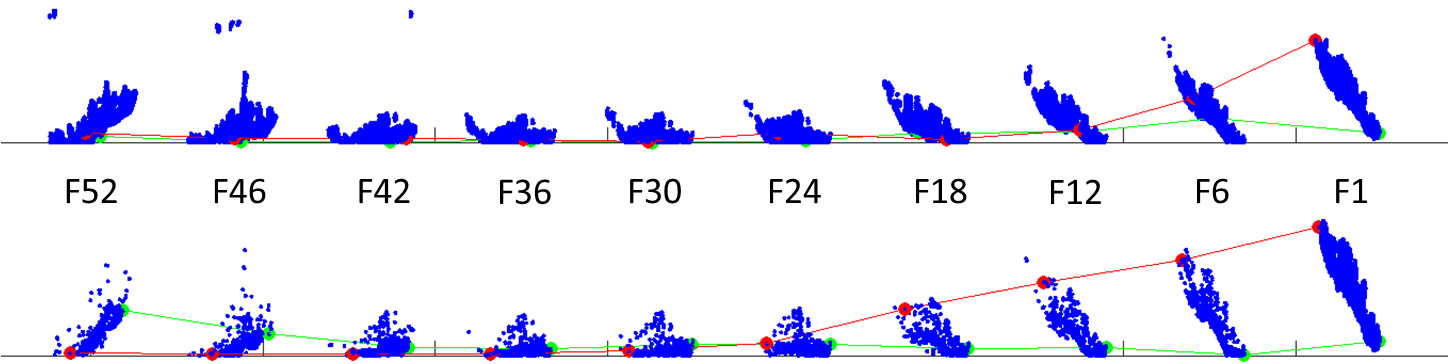}} \vspace{-0.4 cm}\\
    \subfloat[Camera 6]{\includegraphics[width =  0.8\linewidth, height = 0.17\linewidth]{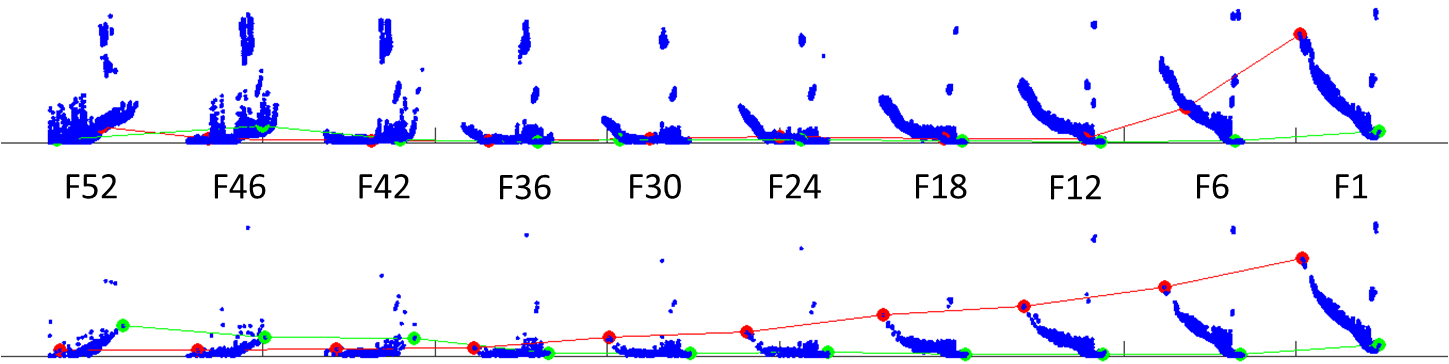}}
    \vspace{-0.2 cm}
    \caption{Top: Raw data; Bottom: Dynamic-Synchronised Graph as the key nodes.}
	\label{fig:meshMorph}
	\vspace{-0.43 cm}
\end{figure}

After scanning, each camera gives a set of time series 3D point clouds, and there are not logical correspondences among them. This prevents us to explain any dynamic features between the frames as the correspondence of points from one frame to the other frame, known as dynamic points, does not exist. Figure~\ref{fig:meshMorph} (the first row of each sub-figure) presents this "lack of correspondence", where we selected two points (highlighted with red and green colors) in the first frame of each camera, and tracked these points in the rest frames using point ids in the acquired point clouds. To be able to have meaningful dynamic features between frames of a camera needed as the key nodes used for the ADGC-LSTM network in Section~\ref{subsec:DSG} , we established the correspondences of points using the Nonrigid Iterative Closest-Farthest Points  (ICFP) scheme \cite{tajdari2022dynamicregistration} which guarantees to find proper corresponded points in a limited number of iterations from a Source mesh ($\mathbb{S}$) to a Target mesh ($\mathbb{T}$). 
In the process of finding correspondences from each point on $\mathbb{T}$ to $\mathbb{S}$, initially, each point on $\mathbb{S}$ may have multiple corresponding points on the $\mathbb{T}$. In this case, we logically select either the closest or the farthest point. In each iteration of the registration process, a boundary  distance ($l$) in Eq(\ref{eq:bound}) is defined as the corresponding distance matrix from $\mathbb{T}$ to $\mathbb{S}$.\vspace{-0.2 cm}
\begin{equation}
l = m + \zeta \sigma
\label{eq:bound}
\vspace{-0.2 cm}
\end{equation} 
where $m$ and $\sigma$ are the mean and standard deviation of the counted distances for the correspondences from $\mathbb{T}$ to $\mathbb{S}$, respectively. $\zeta$ is the probability indicator in \cite{tajdari2022dynamicregistration} regulates the closest-farthest point selection. For a point on $\mathbb{S}$, if a number of points on $\mathbb{T}$ are selected, we consider the point with the largest distance among the selection, if all the distance for the population is greater than the $l$. Otherwise, we select the closest point as the corresponding point to the point on $\mathbb{S}$. 
The ICFP scheme was used to find the available correspondences for all frames captured by a single camera e.g, top-rows of the sub-figures in Figure~\ref{fig:meshMorph}. In the implementation, we use the ICFP to match each consequent pair of frames (e.g. $i^{th}$ frame and $(i+1)^{th}$ frame), starting from the first frame to the last frame, e.g. for 100 captured frames, 99 pairs were used to generate the correspondences matrix. Apparently, not all points in a frame have correspondences in the neighboring frames, as a new frame may not be able to capture all the points captured in the previous frame e.g, from comparing Figure~\ref{fig:meshMorph}(d)-top with Figure~\ref{fig:meshMorph}(d)-bottom after the 42$^{th}$ frame (F42) the density of Dynamic-Synchronised Graph is reduced. Thus, some frames with very low density point-clouds are skipped due to the lack of correspondences.

\vspace{\SpaceReduction}
\subsubsection{ADGC-LSTM Network}
\label{subsubsec:ADGC_LSTM_Network}
In this section, an end-to-end attention enhancement Dynamic-Synchronised Graph Convolution LSTM network (ADGC-LSTM) for points-network-based action behavior recognition, i.e. human walking, is explained. The captured data of the moving object from each camera can be described as a unique class that has considerable overlap(s) with other classes with respect to the configuration of the cameras. All the  ADGC-LSTM networks designed for each Dynamic-Synchronised Graph in this paper also have similar structural characteristics. First, the linear layer and the shared LSTM layer were employed to collect the feature information for each of the graph. Then, the feature information of each graph was fed into the proposed three layers of the ADGC-LSTM as node depiction to consider the spatiotemporal features in the model.

\paragraph{Dynamic-synchronised Graph Model Based on Human feet dynamic points:} Firstly a linear layer and LSTM layer were employed to convert the 3D coordinate of each key node into a high-dimensional feature space regarding the key node-network sequence. The preliminary linear layer maps the 3D coordinates onto a 256-dimensional vector, as the geometric features $P_t$, i.e., $P_{ti}$ defines the geometry feature of key node $i$. As it includes only geometry information, $P_{ti}$ is effective to proceed with the learning process regarding spatial structure features in graph models. The differential feature $V_{ti}$ between two sequential frames, facilitates the dynamic feature understanding used to train the ADGC-LSTM. The sequential group of features is able to  explain a more sparse domain of feature information better, while the differential of the features is more sensitive to the changes of the feature vectors. Thus, the LSTM layer was utilized to avoid having unnecessary sensitivity between the sequential feature groups. Equation~\eqref{eq:principle} presents this proposition.
\begin{equation}
\vspace{\SpaceReduction}
\begin{split}
    E_{ti} &= f_{lstm} \left( concat \left( P_{ti}, V_{ti} \right) \right) \\
    &= f_{lstm} \left( concat \left( P_{ti}, P_{(t-1)i} \right) \right)
\end{split}
\label{eq:principle}
\end{equation}
where $E_{ti}$ is the augmented featured of key node $i$ at time~$t$.

\vspace{\SpaceReduction}
\paragraph{Learning of the ADGC-LSTM:} 
Finally, the global feature $F^{g}_t$ and local features $F^{l}_t$ at each timestamp were converted to scores $o^{g}_t$ and $o^{l}_t$ of each class. According to \eqref{eq:LSTMgates}, the predicted probability of the $i^{th}$ class can be obtained as:\vspace{\SpaceReduction}
\begin{equation}
\vspace{\SpaceReduction}
    \hat{y}_{ti} = \frac{e^{o_{ti}}}{\sum_{j = 1}^{C} e^{o_{ti}}}, i = 1, \cdots, C
\end{equation}
In the training process, taking into account the hidden state of each time interval, the ADGC-LSTM includes short-term dynamics and the loss function with the structure in 
\eqref{eq:training_model}, extracted to the train model as:\vspace{\SpaceReduction}
\begin{equation}
    \begin{split}
        &\!\!\!L = - \sum_{t=1}^{T_3} \sum_{i=1}^{C} y_i \textrm{log}\hat{y}_{ti}^{\textsl{g}} - \sum_{t=1}^{T_3} \sum_{i=1}^{C} y_i \textrm{log}\hat{y}_{ti}^{\textsl{l}} \\
        &\!\!\!\!\!\!\!\!+ \bar{\lambda}\!\sum_{j=1}^{3} \!\sum_{n=1}^{N}\!\left(\!\!1\!-\!\frac{\sum_{t=1}^{T_j} \alpha_{tnj}}{T_j}  \!\!\right)^{2}\!\!\!\!+\! \bar{\beta} \sum_{j=1}^{3} \frac{1}{T_j}\!\sum_{t=1}^{T_j} \!\!\left(\!\sum_{n=1}^{N} \alpha_{tnj}\!\!\right)^2
    \end{split}
    \label{eq:training_model}
\end{equation}
where $y = (y_1, \cdots, y_c)$ is the ground-truth label. $T_j$ denotes the number of time intervals on the $j^{th}$ ADGC-LSTM layer. The third term is considered to emphasize equally to variation of featured points. The final term is to restrict the number of interested nodes. $\bar{\lambda}$ and $\bar{\beta}$ are weight decaying coefficients.

\vspace{\SpaceReduction}
\subsubsection{Implementation of ADGC-LSTM}
\label{subsubsection:Implementation_ADGC_LSTM}
\begin{figure*}[tb]
	\centering
	\includegraphics[width= 0.7\linewidth]{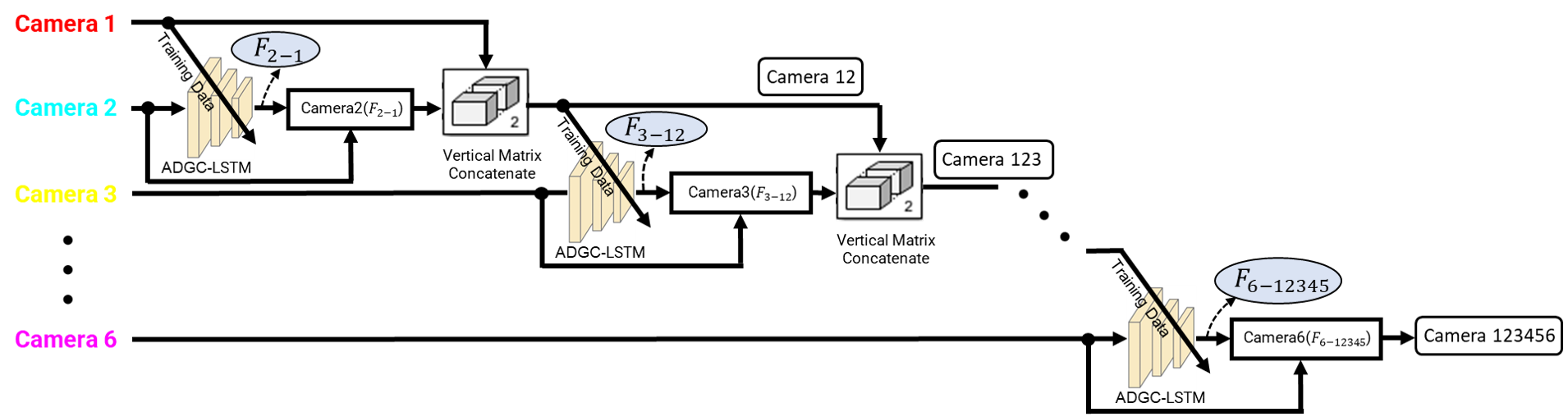}
 \vspace{-0.2 cm}
	\caption{Hierarchical learning-synchronization process.}
	\label{fig:FrameSyncLSTM}
	\vspace{-0.5 cm}
\end{figure*}

To synchronize the captured frames between each pair of cameras, we use the generated Dynamic-Synchronised
Graph of each camera. We use one camera's Dynamic-Synchronised Graph as the supervisor to train our ADGC-LSTM Network and the other one for validation. In this case, we use a hierarchical learning process to have the maximum overlap between cameras with the shown framework in Figure~\ref{fig:FrameSyncLSTM}. In the figure, firstly we synchronize Camera 2 with Camera 1 (where the corresponding frames of Camera 2 to Camera 1 is $F_{2-1}$) and name the overall point cloud as Camera 12 (Camera 1 with synchronized Camera 2). Then synchronize Camera 3 with Camera 12 (with frame set of $F_{3-12}$) and name it as Camera 123. Then we continue with Camera 4, Camera 5, and finally Camera 6, to have all the cameras synchronized based on Camera 1. 

\subsection{Mesh Registration}
\label{subsec:Mesh morphing}

Based on the established correspondences, a cost function based on Tajdari et. al~\cite{tajdari2022feature} is defined for registering meshed at each time step. Tajdari et. al~\cite{tajdari2022feature, tajdari2023non} proposed the non-rigid registration formulation as a combination of distance ($W, D, U$), stiffness ($M, G$), and semi-curvature ($W_c, A_c, B_c$) terms summarised in the following formula\vspace{-0.2 cm}
\begin{align}
E(X) &= \norm{\left[\begin{matrix}
	\alpha M \otimes G\\
	WD\\
	\beta W_cA_c
	\end{matrix}\right]X-\left[\begin{matrix}
	0\\
	WU\\
	\beta W_c B_c
	\end{matrix}\right]}^{2}_{F} \nonumber\\
&= \norm{AX-B}^{2}_{F}
\label{eq:E_final}
\end{align}
where, The sparse matrix $D$ is formed to facilitate the transformation of the source vertices with the individual transformations contained in $X$ via matrix multiplication, and denoted as $D = diag(v_1^T, v_2^T, \dots , v_n^T)$, where $v_i \in \mathbb{S}$ and $i = 1,...,n$, and $n$ is the number of vertices on the $\mathbb{S}$. $W$ is a diagonal matrix consisting of weights $w_i$. $\alpha$ is the stiffness constraint. To regularise the deformation, an additional stiffness term is introduced. Using the Frobenius norm $\norm{.}_F$, the stiffness term penalizes the difference of the transformations of neighboring vertices, through a weighting matrix $G = diag(1, 1, 1, \gamma)$. During the deformation, $\gamma$ is a parameter to stress differences in the skew and rotational part against the translation part of the deformation. The value of $\gamma$ can be specified based on data units and the types of deformation \cite{amberg2007optimal}. The node-arc incidence matrix M (e.g. Dekker \cite{dekker1986mathematical}) of the template mesh topology is employed to convert the stiffness term into the matrix form. As the matrix is fixed for directed graphs, the construction is one row for each edge of the mesh and one column per vertex. To establish the node-arc incidence matrix of the source topology, the indices (i.e. the subscripts) of edges and vertices are addressed, for any edge of $r$ which is connected to vertices $(i, j)$, in $r^{th}$ row of $M$, and the nonzero entries are $M_{ri} = -1$ and $M_{rj} = 1$.


\section{Experiment setup}  \label{sec:Experiment setup}
\subsection{Data-set}  \label{sec:Data-set}

\subsubsection{Our data-set}
Using the proposed 4D scanner and the novel framework, we tried to build an open-access data-set of 4D feet data regarding significant phases of the gait cycle such as initial contact, foot flat, midstance, heel lift, and toe-off. An experiment was designed and approved by the local human research ethical committee. In the experiment, after a brief explanation, participants first read and signed the consent forms. Subjects under 18 had their consent forms signed by their parents/legal guardians. Then each subject was guided to walk through the glass bridge with his/her bare feet twice regarding the left and the right feet, respectively. Both feet of 59 subjects(26 females ($\female$) and 33 males ($\male$)) were scanned while the data of participant 53 was not saved and was excluded from the data-set, resulting in a data-set with 58 subjects. Among them, 55 subjects are right-handed and the rest are left-handed. The age of the population ranges from 6 to 50 years old where the mean age is 24 for females ($\female$) and 26.2 for males ($\male$). Their normal shoe sizes range from 32 to 46 (European sizes, 20-29.3 CM). To be more inclusive and address the diversity of the population, we invited subjects from different countries such as The Netherlands, Belgium, Italy, Spain, Latvia, Slovenia, Swaziland, Turkey, Iran, India, Thai, China, Japan, Costa Rica, Mexico, Cameroon, Nigeria. The anthropometric data of the population can be found in Table~\ref{tab:Ourdata_set}.  
\begin{table}[tb]
	\caption{The anthropometric data.}
	\label{tab:Ourdata_set}
	\centering

\begin{tabular}{p{0.2cm} p{1.1cm} p{1.1cm} p{1.2cm} p{1.2cm} p{1.1cm}}
	\hline\hline
	 Sex & Age & Shoe size & Height & Weight & BMI
	 \\
	\hline
	$\female$	& 24.0$\pm$5.1 & 37.5$\pm$1.5 & 161.1$\pm$9.0 & 55.9$\pm$9.4  & 21.5$\pm$2.6 
	\\
    $\male$   & 26.2$\pm$6.4 & 42.9$\pm$1.6 & 178.8$\pm$8.8 & 73.3$\pm$11.7 & 22.9$\pm$3.0 
    \\
	\hline\hline
\end{tabular}



	\vspace{-0.43 cm}
\end{table}

\vspace{\SpaceReduction}
\subsubsection{Data-set for registration}
\label{subsec:RegistrationFoot}
In the experiment, both the right and the left feet shapes in data-set number 25 in the SHREC'14 data-set \cite{Pickup2014} were selected as the source surface. Before the experiment, the meshes of both feet were pre-processed for a more uniform mesh using ACVD, a freely available software provided by Valette et al. \cite{valette2008generic}. The acquired two meshes, each has 5000 vertices, were used as the inputs of the experiment as the source meshes for the nonrigid registration regarding the left and the right foot, respectively.

\vspace{-0 cm}
\subsection{Methods for comparison}
\label{subsec:comparison}

We compare the  proposed methods in the framework with the following methods with similar state-of-the-art~\cite{guo2019attention}:
\begin{itemize}
    \item ARIMA \cite{williams2003modeling}: Auto-Regressive Integrated Moving Average method is one of the well-known methods to anticipate the future values in a time sequential data-set.
    \item VAR \cite{zivot2006vector}: Vector Auto-Regressive finds the pairwise connectivity between time-sequential data-sets.
    \item LSTM \cite{hochreiter1997long}: Long-Short Term Memory network, is a variant of RNN network.
    \item GRU \cite{chung2014empirical}: Gated Recurrent Unit network, is a specific RNN network.
    \item STGCN \cite{yan2018spatial}: A Spatial-Temporal Graph Convolution model is developed based on automatic learning of both the spatial and temporal patterns.
    \item GeoMAN \cite{liang2018geoman}: A multi-level attention-based RNN model aimed for the geo-sensory time sequential anticipation problem.
\end{itemize}

Root mean square error (RMSE) of the geometry based on closest points is used as the metric.

\color{black}
\subsection{ADGC-LSTM parameters' configuration}

In the experiments, a fixed length of $T = 40$ is used in \eqref{eq:training_model} from
each graph sequence as the input. Regarding the ADGC-LSTM, we assumed the neighbor set of each node includes only nodes directly connected with itself. 
Regarding a fair comparison with ST-GCN~\cite{yan2018spatial}, the graph labeling function in ADGC-LSTM divides the neighbor set into $K = 3$ subsets according to~\cite{yan2018spatial}.
In the training process, the Adam optimizer~\cite{kingma2014adam} is employed to optimize the network. The parameters of $\bar{\lambda}$ and $\bar{\beta}$ are set to 0.01 and 0.001, respectively. We set the initial learning rate to 0.0005 which is reduced in every 15 epochs by multiplying 0.1 to the learning rate. In addition, we used the same $\zeta = 1.7$ in \eqref{eq:bound} as \cite{tajdari2022dynamicregistration}. Regarding the used mesh registration method in Section~\ref{subsec:Mesh morphing}, we use the same parameter values in ~\cite{tajdari2022feature} regarding \eqref{eq:E_final}.
\color{black}

\section{Results}
\label{sec:Results}

\subsection{Motion synchronisation}

To evaluate the effectiveness of the synchronization, we developed a K-fold-like scheme where: 1) we used the mean Closest Points Geometry Distance (CPGD) \cite{amberg2007optimal} values between adjacent point clouds as the metric and 2) for each camera, we compared its synchronized scans to the merged results of other 5 cameras at each timestamp. That is, in the $i^{th}$ frame and after synchronization with each of the aforementioned methods in Section~\ref{subsec:comparison}, we exclude the $j^{th}$ camera points from the complete foot and calculate the CPGD of the camera $j^{th}$ points with the remaining points. We repeat this process for all other cameras and the average values of errors are presented in Table~\ref{tab:ComparisionResults}. According to the table, our proposed method outperforms all the other methods for the both the left and right feet in the data-sets.

\begin{figure}[tb]
		\centering
    \subfloat[ARIMA]{\includegraphics[width = 0.9\linewidth]{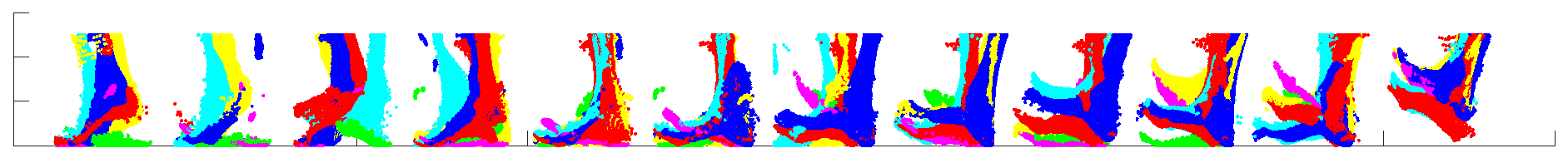}} \vspace{-0.4 cm}\\
    
    \subfloat[VAR]{\includegraphics[width =  0.9\linewidth]{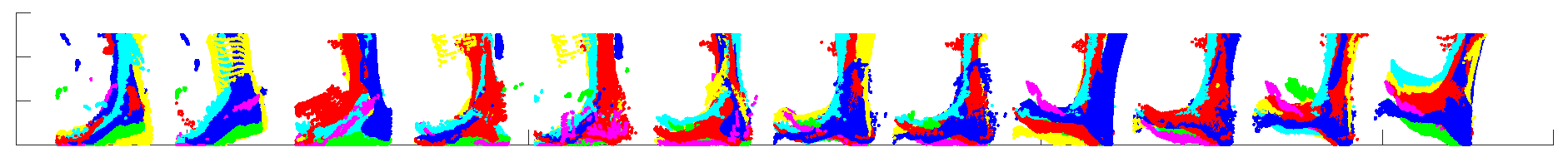}} \vspace{-0.4 cm}\\
    \subfloat[LSTM]{\includegraphics[width =  0.9\linewidth]{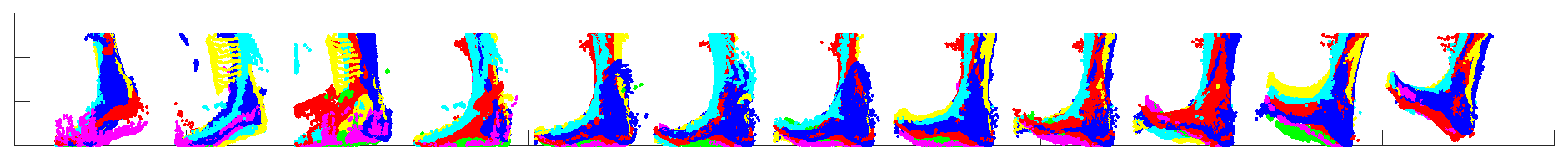}} \vspace{-0.4 cm}\\
    
    \subfloat[GRU]{\includegraphics[width =  0.9\linewidth]{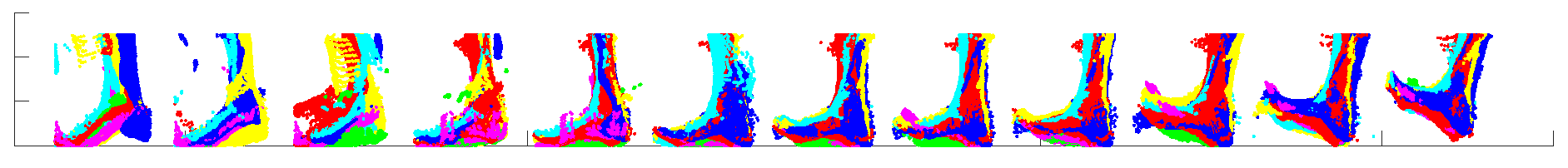}} \vspace{-0.4 cm}\\
    \subfloat[STGCN]{\includegraphics[width =  0.9\linewidth]{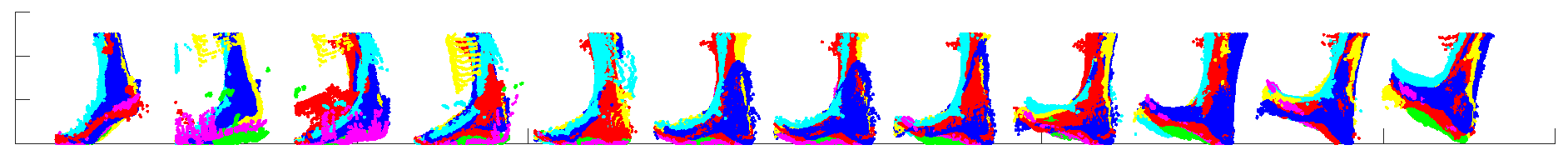}} \vspace{-0.4 cm}\\
    \subfloat[GeoMAN]{\includegraphics[width =  0.9\linewidth]{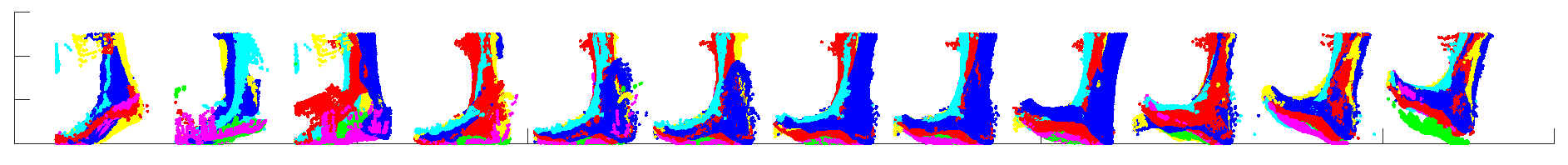}} \vspace{-0.4 cm}\\
    \subfloat[Our]{\includegraphics[width =  0.9\linewidth]{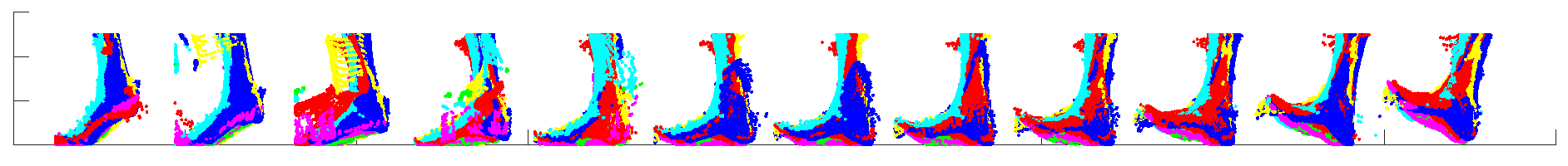}}
    \vspace{-0.3 cm}
    \caption{Results of time synchronisation with different methods.}
	\label{fig:SyncedResults}
	\vspace{-0.43 cm}
\end{figure}

According to Table~\ref{tab:ComparisionResults} and Figure~\ref{fig:SyncedResults}, one can be seen is that generally the output of the non-deep learning methods e.g., ARIMA and VAR, demonstrate a higher error than the deep learning methods e.g., LSTM, GRU, STGCN, GeoMAN. This is investigated numerically and the results are presented in Table~\ref{tab:ComparisionResults}, which shows that the deep-learning methods could averagely improve the performance for about 80\% in terms of PI, revealing the limited abilities of the non-deep-learning methods to tackle non-linearity and complexity in time series analysis. Among the deep-learning methods, the models that simultaneously consider temporal and spatial correlations, e.g. STGCN, GeoMAN, and the proposed method, outperform other deep-learning-based methods including LSTM and GRU for about 11\% in terms of PI. Where, GeoMAN slightly outperforms STGCN in terms of PI, defining that the multi-level attention mechanisms employed in GeoMAN enhance finding the correlation among dynamic features of the feet. Our ADGC-LSTM network, achieved better results than other included  state-of-the-art methods, confirming the performance of the proposed method in describing spatial-temporal features of the walking foot.
\begin{table}[tb]
	\caption{RMSE results of the comparison based on Closest Points Geometry Distance (CPGD), and Percent of Improvement (PI) comparing to the raw data, for the left and right foot.}
	\label{tab:ComparisionResults}
	\centering
	\vspace{-0.3 cm}

\begin{tabular}{p{1cm} p{0.02cm} p{1cm} p{1cm} p{0.02cm} p{1cm} p{1cm} }
	\hline\hline
	\multirow{2}{*}{Method} & & \multicolumn{2}{c}{Left foot} & & \multicolumn{2}{c}{Right foot}\\ \cline{3-4} \cline{6-7}
	
	& & CPGD (cm) & PI(\%) & & CPGD (cm) & PI(\%)  \\
	\hline\hline
	raw data  & & 7.01 & -- & & 7.35 & -- \\
	\hline
	ARIMA  & & 6.81 & 2.8  & & 5.95 & 19.1 \\
	VAR    & & 3.24 & 53.8 & & 4.22 & 42.6 \\
	LSTM   & & 1.62 & 76.9 & & 1.58 & 78.5 \\
	GRU    & & 1.71 & 75.6 & & 1.59 & 78.3 \\
	STGCN  & & 1.21 & 82.7 & & 1.14 & 84.4 \\
	GeoMAN & & 1.13 & 83.8 & & 1.03 & 85.9 \\
	Our   & & \textbf{0.64} & \textbf{90.8} & & \textbf{0.72} & \textbf{90.2} \\
	\hline\hline
\end{tabular}
	\vspace{-0.43 cm}
\end{table}

\subsection{Establishing registration-based data-set}
\vspace{-0.2 cm}

Through the explained method in Section~\ref{subsec:Mesh morphing}, we register the synchronised frames in Section~\ref{subsec:RegistrationFoot} at each time step to establish a mesh morphed 3D geometry as Figure~\ref{fig:Our_Data_set}.
In this regard, we can track not only the geometry of any point but also the dynamic features of the point such as velocity, and acceleration. In addition, we can compare the geometry or dynamic features of any points among all captured feet shapes. To this end, we numerically investigated the deformation variation of a few well-known foot dimensions in Table~\ref{tab:DimensionComparision}. The dimensions are length ($L_f$), width ($W_f$), and ball width ($BW_f$) according to~\cite{Tang2017FootDesign} used in \cite{tajdari2022dynamicregistration}, and their variations are $\Delta L_f$, $\Delta W_f$, and $\Delta BW_f$. Where the operator $\Delta$ defines the differences between the maximum and minimum value of the dimension for a participant during walking.
\begin{table}[tb]
	\caption{Foot dimensions variation results.}
	\label{tab:DimensionComparision}
	\centering
        \vspace{-0.3 cm}

\begin{tabular}{p{2.4cm} p{1.8 cm} p{2.3 cm} }
	\hline\hline
	Parameter & Left foot & Right foot\\
	\hline
	$\Delta L_f$ (cm) & 0.71$\pm$0.60 & 0.69$\pm$0.62\\
	$\Delta W_f$ (cm) & 1.1$\pm$0.9 & 1.0$\pm$0.9\\
	$\Delta BW_f$ (cm) & 1.2$\pm$0.9 & 1.1$\pm$1.0\\
	\hline\hline
\end{tabular}
	\vspace{-0.43 cm}
\end{table}
By calculating the average foot length ($L_{ave}$) of all the feet in our data-set (both left and right feet) as 24.3 cm, we can see from Table~\ref{tab:DimensionComparision} that the variation of $L_f$ is about 3\% of $L_{ave}$, and $W_f$ and $BW_f$ are about 5\% of $L_{ave}$, which are a considerable variation and highlights the importance of 4D scanning, and 4D studying of human actions.

\vspace{\SpaceReduction}
\section{Conclusion} \vspace{-0.2 cm}
In this paper, we proposed a generic framework to synchronize and register asynchronously captured point clouds of a moving and deforming object, namely the human foot, through a novel ADGC-LSTM-based network and a non-rigid registration algorithm. We implemented the framework on the data captured from a novel 4D foot scanner to acquire the first 4D open-access feet data-set with the focuses on 1) finding the dynamic connectivity among 3D scans captured at different timestamps of each camera in terms of dynamic feature synchronization and 2) extracting meaningful dynamic features from the combined views of multiple cameras for estimating the amplitude of the deformation. Experiment results show that our method improved the synchronization process on average by about 30\% compared to other state-of-the-art methods. Meanwhile, the quality of the acquired 4D scan was comparatively high regarding the deformation of each part of the foot, and such information can be useful in different applications, e.g. footwear design.  
Further developments include establishing a 4D Statistical Shape Model (SSM) of human foot as a tool to study the gait and foot deformations.
Also, due to inconsistency in capturing speeds of different cameras, there are differences in the resolutions of the frames, which might be improved by using temporal super-resolution repetitive motion methods.

{\small
\bibliographystyle{ieee_fullname}
\bibliography{mybibfile}
}

\end{document}